\documentclass[conference]{IEEEtran}

\usepackage[super]{nth}

\usepackage{graphicx}
\usepackage{multirow}
\usepackage{amsmath}
\usepackage{ulem}
\usepackage{comment}

\usepackage{xcolor}

\ifCLASSINFOpdf
\else
\fi

\begin{document}
%
\title{Has Your FaaS Application Been Decommissioned Yet? - A Case Study on the Idle Timeout in Function as a Service Infrastructure}

\author{
\IEEEauthorblockN{Kim Long Ngo}
\IEEEauthorblockA{
York University\\
Toronto, Canada\\
kimlngo@yorku.ca}
\and
\IEEEauthorblockN{Joydeep Mukherjee}
\IEEEauthorblockA{
California Polytechnic State University\\
CA, United States\\
jmukherj@calpoly.edu}
\and
\IEEEauthorblockN{Zhen Ming (Jack) Jiang}
\IEEEauthorblockA{
York University\\
Toronto, Canada\\
zmjiang@cse.yorku.ca}
\and
\IEEEauthorblockN{Marin Litoiu}
\IEEEauthorblockA{
York University\\
Toronto, Canada\\
mlitoiu@yorku.ca}
}

\maketitle

\begin{abstract}
Function as a Service (FaaS) is a new cloud technology with automated resource management. Different from traditional cloud computing, each FaaS cloud function can only run a fixed period of time before being decommissioned. Furthermore, FaaS cloud providers often update their platforms (e.g., idle timeout). These changes and their associated impact are not transparent and could potentially impact the execution of the cloud functions. Hence, in this paper, we develop a methodology to characterize the cloud function idle timeout which is the duration a FaaS cloud provider keeps a cloud function instance alive without serving active traffic. Our study was conducted on three popular FaaS platforms, namely AWS Lambda, IBM and Azure Cloud Function. Moreover, we also report how long a cloud function instance can be kept alive when a user regularly polls the instance. Experimental results show that the idle timeout period has evolved from 01/2020 till 01/2022.
\end{abstract}


%
\IEEEpeerreviewmaketitle

\section{Introduction}
\label{sec: intro}
The software industry has evolved rapidly with the transformation from monolithic architecture operating on Virtual Machines (VM) to Microservices running on light-weight containers such as Docker. To relieve the software engineers from operational tasks such as resource management, a new technology called Function as a Service (FaaS) was popularized since 2014. FaaS is an event-driven cloud platform where software engineers can focus on business logic and leave the infrastructure management to cloud providers~\cite{amazon_building_apps_with_serverless}.

The automated resource management feature brings both advantages and challenges to cloud subscribers. On one hand, software engineers can be relieved from resource management tasks. On the other hand, they do not have the full control over the infrastructure. Without access to traditional technologies like VMs or Containers, the engineers may not know when their cloud function instances are recycled. When a cloud function is first invoked, cloud platform needs to start up a container for execution. This process introduces additional delay in function execution and is referred as cold start. Cold start can occur when a cloud function first started or a cloud instance is left idle beyond a certain time (\textit{idle timeout}) or while scaling up the application. Since cold start introduces execution delays, the more cold start an application experiences, the more overhead is added to the overall response time. As a consequence, knowing when a cloud function instance is decommissioned will help engineers to better design their software systems.

To make FaaS an abstract platform, cloud providers do not specify their underlying implementation including how long a cloud function can be idle. However, researchers have shown the importance of knowing FaaS runtime properties and partially characterized the FaaS function instance's idle timeout \cite{lloyd_factors_influencing_microservice_perf}, \cite{lloyd_keep_alive}, \cite{maissen_faasdom}. Industry practitioners also studied this aspect and reported their experimental results on the web blogs \cite{aws_lambda_terminates_instance}, \cite{shilkov_cold_start_after_10_mins}, \cite{octo_cold_warm_start_aws}. These studies are helpful to software engineers since they provide in-depth knowledge about FaaS runtime. Moreover, our experimental results show that FaaS providers regularly update their implementations. For example, our study between 01/2020 and 01/2022 detected that AWS Lambda cloud function instance idle timeout has been reduced from 10 to 5 minutes and this was not communicated.

In this paper, we characterize FaaS instance idle timeout by presenting a methodology to measure this duration up to minute-accuracy. To keep the cloud function instance from decommissioning, a common workaround is the keep-alive technique which is to poll the cloud function at a regular interval to preserve the infrastructure~\cite{lloyd_keep_alive}. We study the impact of applying the keep-alive technique on the cloud function idle timeout. Furthermore, we also measure the cloud function's instance idle timeout at different times (i.e., checkpoints) from 01/2020 to 01/2022 and report the evolutionary changes. Our experiments were carried out on three popular cloud platforms, namely AWS Lambda (AWS), IBM Cloud Function (IBM) and Azure Cloud Function (Azure). The contributions of our research are:

\begin{itemize}
    \item Presenting a method to measure the cloud function idle timeout with minute-accuracy by extending the state of the art method~\cite{lloyd_factors_influencing_microservice_perf}.
    
    \item Measuring the cloud function instance maximum idle timeout when the keep-alive technique is used and based on this finding, we advise software engineers in which use case they should use the keep-alive technique.
    
    \item Characterizing the cloud function instance idle timeout changes over a period of two years (01/2020 - 01/2022).
\end{itemize}

With our methodology, software engineers can reproduce the experiments to measure the up-to-date cloud function idle timeout or design similar approaches to examine other FaaS' characteristics.

The rest of the paper is organized as follow: Section \ref{sec: background and related work} introduces the background and related work. Section \ref{sec: methodology} discusses our methodology to measure cloud function instance idle timeout. Section \ref{sec: experimental setup} presents the experimental setups used in our studies. Section \ref{sec: evaluation results} reports the evaluation results. Section \ref{sec: summary and lessons learnt} summarizes our findings and highlights the lessons learnt. Finally, Section \ref{sec: conclusion} concludes our research work.

\section{Background and Related Work}
\label{sec: background and related work}
In this section, we first provide an overview of the FaaS in Section \ref{sec:background}. Then we discuss about the related work in Section \ref{sec:related works}.

\subsection{Background}
\label{sec:background}
FaaS is a new cloud technology that automatically manages the computing resources. When traffic first arrives to a cloud function, FaaS platform will provision, deploy the source code and initialize the cloud function instance for execution~\cite{amazon_container_reuse}. This process usually lasts between 300 ms to 24 seconds thus taking the first request longer to completed which denote as \textit{cold start}, the remainder of execution is faster and denoted as \textit{warm start}~\cite{manner_cold_start_factors}. FaaS charges the user based on the cloud function's execution duration hence it is essential to prevent a cloud function from running infinitely. To achieve this, all cloud platforms impose an \textbf{execution timeout} which is the maximum duration a cloud function is allowed to execute. Beyond this period, the platform will stop the execution and return a failed response. Cloud providers allow software engineer to configure their cloud function's execution timeout according to their requirement. Table \ref{table: characteristics and idle timeouts} summarizes the up-to-date documented characteristics including maximum execution timeout obtained from the cloud providers.

To minimize the overhead caused by cold start, after serving the current traffic, cloud platforms will retain the infrastructure for a short time to accommodate future requests. If there is no traffic within this idle period, cloud platforms will decommission and return the computing resources for other purposes. We refer this idle period as cloud function \textbf{idle timeout}. To our knowledge, cloud providers do not officially document how long the idle timeout lasts.

\subsection{Related Work}
\label{sec:related works}
FaaS infrastructure retention has been studied when researchers explored FaaS' characteristics. Lloyd et al. \cite{lloyd_factors_influencing_microservice_perf} in their study about serverless computing reported that all FaaS VMs and containers were removed after being left idle for 40 minutes and by using the keep-alive technique at 5-minute interval, host VMs were recycled every 4 hours. Maissen et al.~\cite{maissen_faasdom} described that on AWS and IBM, it took around 10 minutes of no activity for cloud function instance to be recycled. This period for Azure cloud function was higher at 20 minutes.

We note that previous research results have been outdated since cloud providers regularly update their implementation. Our study extends the previous state-of-the-art~\cite{lloyd_factors_influencing_microservice_perf} and presents a method to quantify the cloud function idle timeout. In addition, we move one step further to summarize the evolutionary changes from 01/2020 to 01/2022. FaaS instance idle timeout and corresponding cold start can impact the FaaS' performance and software system, and therefore they can be used as an input in formal models or performance models where timing or Service Level Agreements (SLA) are a strict requirement. Our source code is published for reproduction purposes~\cite{zenodo_repo}.

\section{Methodology}
\label{sec: methodology}
In this section, we describe our methodology to measure function instance idle timeout. We discuss our measurement methodology in Section \ref{subsec:methodology function idle timeout measurement} followed by how we uniquely identify a function instance in Section \ref{subsec:Function Instance Identification}. In Section \ref{subsec:Function Instance Keep-Alive Idle Timeout}, we outline how we measure the function instance idle timeout when the keep-alive technique is used.

\subsection{Function Idle Timeout Measurement}
\label{subsec:methodology function idle timeout measurement}
To measure a platform's FaaS idle timeout, we first deployed our testing function (discussed in Section \ref{subsec: testing function}). Next, we invoked the function periodically with reducing interval to detect the idle timeout. We started with a 20-minute interval because we noticed that function instances would be decommissioned if they were left idle for this duration. Subsequently, we decreased the interval by one minute, ran the test for five hours and checked if the requests were served by the same instance. If all the requests were served by different function instances, we would then proceed to the next reduced interval. The experiments continued until we detected an interval that made two consecutive requests served by the same instance. We noted this idle timeout as $x$ minutes and repeated the experiment between $x$ and $(x+1)$ minutes for five hours to ensure consistent results. To detect the evolutionary changes cloud providers implemented on function instance idle timeout, we conducted the regular experiments over the following extended period [12/09 - 01/10/2020], [27/03 - 13/05/2021], [19/07 - 27/07/2021] and [08/01 - 15/01/2022].

\subsection{Function Instance Identification}
\label{subsec:Function Instance Identification}
To measure the idle timeout of a function instance, we measure the longest duration that a container instance can be left idle before it is decommissioned. This is equivalent to measuring the longest duration that makes two consecutive requests served by the same instance. Hence, uniquely identifying a cloud function instance is necessary. To achieve this goal, we used the following approach: 
\begin{itemize}

\item For AWS Lambda, we obtained the \textit{``logStreamName''} which is tied to the function executing instance and extracted from the execution context.

\item For Azure Cloud Function, we queried the Monitoring - Log to retrieve requests' \textit{``customDimensions''} which contained the executing instance identifier.

\item For IBM Cloud Function, we extended the self-generate methodology presented by Lloyd et al.~\cite{lloyd_factors_influencing_microservice_perf} to create a universally unique identifier (UUID) for the function instance. When a cloud function is invoked, it checks the local file \textit{``/tmp/host.txt''} for the UUID. If this file does not exist, it means this is a new instance and a UUID is created and stored in this file. To ensure thread safety, we implemented the synchronization with double lock mechanism while creating and storing the UUID. When the \textit{``/tmp/host.txt'}' exists, subsequent execution can access the file and retrieve the UUID without acquiring the lock to prevent performance degradation.
\end{itemize}

\subsection{Function Instance Keep-Alive Idle Timeout}
\label{subsec:Function Instance Keep-Alive Idle Timeout}
Once we determine the function instance idle timeout as $x$ minutes, we invoke the function with inter-arrival request rates less than x minutes and measure the time duration the function instance can be kept-alive. The experimental results were to answer the question if the keep-alive technique can be used as a work around to resolve the cold start issue. Our experiments were carried out between [19/07 - 27/07/2021] and [08/01 - 15/01/2022].

\section{Experimental Setup}
\label{sec: experimental setup}
In this section, we discuss how we setup our cloud functions on different cloud platforms.

\subsection{Cloud Providers}
Based on Eismann et al.\cite{eismann_serverless_use_case} FaaS' use case study, we chose to evaluate FaaS on three most popular cloud providers, namely AWS Lambda, Microsoft Azure Function and IBM Cloud Function because they occupied the majority of FaaS use cases (80\%, 10\% and 7\% respectively). We did not study other providers like Google Cloud Function, as they only occupy a small use case percentage (3\%).

\subsection{Runtime}
We focused on Java in this paper because Java and Node.js are known to be the most popular studied languages in FaaS software industry research by Scheuner et al.~\cite{scheuner_multivocal_literature}. Among different versions of Java, we used Java version 8 since it is the commonly supported version on all cloud platforms.

\subsection{Testing Function}
\label{subsec: testing function}
We followed Spillner et al. \cite{spillner_prospect} and Manner et al. \cite{manner_cold_start_factors} to setup a testing function which computes the $n^{th}$ value of the Fibonacci (fib) sequence recursively. We also used $fib(38)$ as the processing workload.

\begin{figure}[htbp]
\centerline{\includegraphics[width=0.4\textwidth]{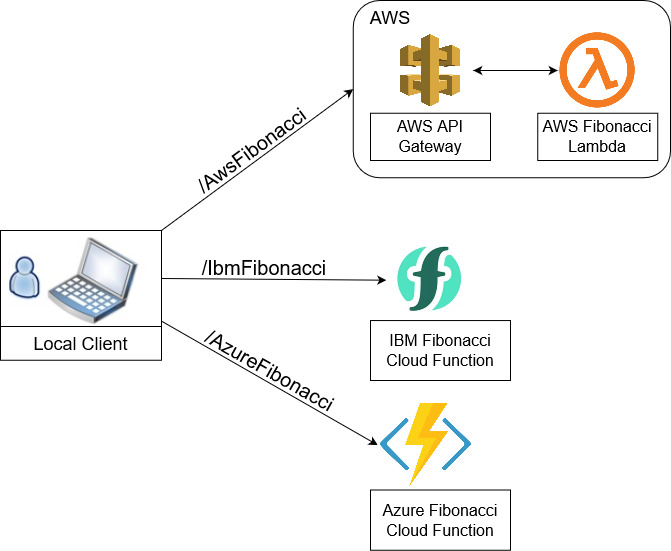}}
\caption{Experimental Setup.}
\label{fig:experimental setup}
\end{figure}

For AWS Lambda, we integrated the cloud function with  API Gateway to enable HTTPS communication. IBM and Azure Cloud Functions had this capability supported by default hence no additional implementation was required. The experimental system is shown in Fig. \ref{fig:experimental setup}.

We configured the AWS Lambda and IBM Cloud Function instances to have 512MB memory and 15-second execution timeout as experimental results showed that these configurations were appropriate for the Fibonacci task. Azure Cloud Platform did not have memory configuration options hence we deployed our Azure Fibonacci cloud function on a single Function App running on Linux Operating System with 15-second execution timeout. These deployment settings help to avoid resource-sharing which can interfere with experimental results. In addition, for comparison purpose, we also created a Hello-World cloud function and deployed with the same cloud instance configurations. The Hello-World function returns a welcome message upon invocation and processes no business logic. The cold and warm start average response times of the two experimental cloud functions are shown in Table \ref{table: cold-warm-start response time}.

\section{Evaluation Results}
\label{sec: evaluation results}
Here we present our experimental results.

\begin{table}[h!]
\fontsize{10}{11}\selectfont
\centering
\renewcommand{\arraystretch}{1.15}
\begin{tabular}{|l|c|c|c| } 
 \hline
 \textbf{Characteristics} & \textbf{AWS} & \textbf{IBM} & \textbf{Azure} \\
 \hline
 Max Execution Timeout (mins) & 15 & 10 & 10 \\
 \hline
 Max Allocated Memory (GB) & 10 & 2 & 1.5 \\
 \hline
 Billing Granularity (ms) & 1 & 100 & 1 \\ 
 \hline
 \multicolumn{4}{|l|}{\textbf{Cloud Function Idle Timeout}} \\
 \hline
 Cloud Function & \multirow{2}{*}{5} & \multirow{2}{*}{10} & \multirow{2}{*}{12} \\ 
 Idle Timeout (mins) & & & \\
 \hline
 Maximum Idle Timeout & \multirow{2}{*}{145} & \multirow{2}{*}{336} & \multirow{2}{*}{2675} \\
 Using Keep-Alive (mins) & & & \\
 \hline
 \nth{90}-Percentile Idle Timeout & \multirow{2}{*}{140} & \multirow{2}{*}{138} & \multirow{2}{*}{1639} \\
 Using Keep-Alive (mins) & & & \\
 \hline
 \multicolumn{4}{|l|}{\textbf{Evolutionary Cloud Function Idle Timeout Changes}} \\
 \hline
 01 - 02/2020 [acc. Maissen] & 10 & 10 & 20 \\ 
 \hline
 09 - 10/2020 & 10 & 10 & 14 \\
 \hline
 03 - 05/2021 & \multirow{2}{*}{5} & \multirow{2}{*}{10} & \multirow{2}{*}{12} \\
 \cline{1-1}
 07/2021 - 01/2022 & & &\\
 \hline
\end{tabular}
\renewcommand{\arraystretch}{1}
\caption{Summary of FaaS Characteristics (Documented) and Cloud Function Idle Timeout And Evolutionary Changes (Undocumented)}
\label{table: characteristics and idle timeouts}
\end{table}

\begin{table}[h!]
\fontsize{10}{11}\selectfont
\centering
\begin{tabular}{|l|c|c|c| } 
 \hline
 \textbf{Response Time} & \textbf{AWS} & \textbf{IBM} & \textbf{Azure} \\
 \hline
 Fibonacci (Cold start) & 1,161 & 3,169 & 2,825 \\ 
 \hline
 Fibonacci (Warm start) & 778 & 695 & 628 \\
 \hline
 Hello-World (Cold start) & 698 & 1,495 & 2,663 \\ 
 \hline
 Hello-World (Warm start) & 79 & 169 & 81 \\
 \hline
\end{tabular}
\caption{Average Response Time (in ms) during Cold and Warm Start.}
\label{table: cold-warm-start response time}
\end{table}

\subsection{Function Instance Idle Timeout}
Table \ref{table: characteristics and idle timeouts} shows the un-documented cloud function instance idle timeout, the maximum and \nth{90}-percentile idle timeout when the keep-alive technique is used. The results show that when being left idle, a cloud function instance can be retained at most \textbf{5 minutes}, \textbf{10 minutes} and \textbf{12 minutes} for AWS Lambda, IBM and Azure Cloud Function, respectively. Beyond this period, cloud platforms will reclaim the function instance and subsequent requests will encounter cold start.

Upon completing the cloud function instance timeout on AWS Lambda and IBM Cloud Function with 512MB allocated memory, we re-examined the experiments on these cloud platforms with 1024MB memory to understand if there is a relation between this timeout and the allocated memory. This additional experiment is not applicable to Azure Cloud Function as this platform does not support memory configuration. The results show no difference compared to results in Table \ref{table: characteristics and idle timeouts} hence we conclude that there is no relation between a cloud instance idle timeout and allocated memory for AWS Lambda and IBM Cloud Function.

\subsection{Maximum Function Instance Idle Timeout When Using Keep-Alive Technique}
Based on the maximum idle timeout presented above, we configured our local timer clients to poll every 5, 10 and 12 minutes for AWS Lambda, IBM and Azure cloud functions, respectively, to examine how long the function instances can be re-used if they are kept warm.

The results show that an AWS Lambda cloud instance can be retained at most \textbf{145 minutes}. The \nth{90}-percentile for the duration of service is \textbf{140 minutes}. The small gap between the maximum and \nth{90}-percentile implies that this cloud platform adopts a static recycling algorithm and most of the function instance will be recycled after 140 minutes.

In contrast, IBM cloud function instance can serve a periodic traffic continuously at most \textbf{336 minutes}. Nevertheless, the \nth{90}-percentile value drops to \textbf{138 minutes}. The enormous gap between the maximum and \nth{90}-percentile may infer that IBM cloud platform occasionally kept the cloud instance longer than usual, however, majority of the function instances will be decommissioned after around 138 minutes.

Microsoft Azure cloud function exhibited an interesting behavior. When tested with the polling period between 6 to 12 minutes, a function instance can be used at most \textbf{20 minutes}. However, when we reduced the polling period to 5 minutes, we observed the function infrastructure was retained up to \textbf{44 hours 30 minutes}. This behavior may induce that Azure cloud function considers the 5-minute interval as a frequent invocation pattern hence the platform keeps the function instance warm to serve the traffic. On the contrary, a longer period between 6 to 12 minutes may not be frequent and hence the function instance is recycled after maximum 20 minutes.

\subsection{Function Instance Idle Timeout Evolutionary Changes}
Applying our methodology at different checkpoints from 09/2020 to 01/2022, we discovered that cloud function's instance idle timeouts have implicitly changed. We first noted the previous study's result reported by Maissen et al. \cite{maissen_faasdom} which were established during the [01 - 02/2020] and then we conducted four experiments during: [09 - 10/2020], [03 - 05/2021], 07/2021 and 01/2022. The measurement results summarized in Table \ref{table: characteristics and idle timeouts} shows that IBM Cloud Function did not change the function idle timeout since 01/2020 to 01/2022. AWS Lambda, however, reduced this duration by 50\% from 10 to 5 minutes some time between 10/2020 and 03/2021. Microsoft Azure cloud platform also shortened the instance's idle timeout by nearly 50\% over the time. These changes were neither documented nor communicated to software engineers.

\section{Summary and Lessons Learnt}
\label{sec: summary and lessons learnt}
FaaS is designed to be ephemeral; cloud providers will decommission the cloud function instance if it is left idle for 5 to 12 minutes depending on the platform. Software engineers can expand the idle timeout period up to hours by using the keep-alive technique. Nevertheless, all cloud providers will eventually reclaim their cloud function instances regardless of active traffic. The decision to use the keep-alive technique depends on the application use case. For instance, if the application only serves the traffic at certain time, software engineers may not need the keep-alive technique. If the traffic pattern is random and there is an enormous difference between cold and warm start like the Hello-World example, it might be helpful to apply the keep-alive technique to reduce performance overhead. Through our experimental results and Maissen et al.~\cite{maissen_faasdom} study, we observed that cloud function idle timeout has been reduced over time. These changes are not documented by cloud providers but only reported by research studies \cite{lloyd_factors_influencing_microservice_perf}, \cite{lloyd_keep_alive}, \cite{maissen_faasdom} or by industrial web blogs \cite{aws_lambda_terminates_instance}, \cite{shilkov_cold_start_after_10_mins}, \cite{octo_cold_warm_start_aws}.

By knowing the cloud function instance idle timeout of each cloud platform, software engineers can evaluate a system's performance based on the traffic pattern. If traffic requests arrive within the idle timeout, cold starts are avoided and hence the requests are served with low latency. In contrast, if the inter-arrival request rates fall beyond the cloud function idle timeout, cold starts occur and thus function's performance degrades. In cases where performance guarantees are required, software engineers may consider using keep-alive techniques or employ a cloud provider's solution to keep the instance warm. Examples are Provisioned Concurrency from AWS Lambda and Dedicated or Enterprise subscription plans from Azure. These technical solutions incur additional cost but cloud function instances are kept warm and thus resolve the cold start issues.

\section{Conclusion}
\label{sec: conclusion}
In this paper, we presented a method to quantify the FaaS instance's idle timeout up to minute-accuracy. Furthermore, we also examined this idle timeout maximum value when the keep-alive technique is used and reported the evolutionary changes that we detected through 01/2020 - 01/2022.

In the future, we plan to evaluate actual software systems with live traffic patterns using our knowledge of the cloud function idle timeout. This is to determine if the reduction in cloud function idle timeout will degrade the system's performance. In addition, we also plan to compare the operating cost between the two approaches: (1) using the keep-alive technique as discussed in this paper and (2) using pre-warmed instances provided by cloud providers, e.g., provisioned concurrency in AWS Lambda. The findings can provide software engineers with thorough understanding about the performance and cost impact to their systems when the cloud function idle timeout is changed on cloud platforms.


%

\end{document}